\documentclass{iopconfser}
\usepackage[dvips]{epsfig}
\usepackage{geometry}
\usepackage{hyperref}
\usepackage{pdfpages}
\usepackage{subfigure}
\usepackage{caption}
\usepackage{subcaption}
\begin{document}

\title{High Reynolds number trends of centerline mean velocity and normal stress in pipe flow }

\author{Hassan Nagib$^{1}$, Lorenzo Lazzarini$^{2}$, Gabriele Bellani$^{2}$ and Alessandro Talamelli$^{2}$}

\affil{$^1$Mechanical, Materials and Aerospace Engineering Department, Illinois Tech (IIT), Chicago, USA}
\vspace{0.3 cm}
\affil{$^2$Department of Industrial Engineering, University of Bologna, Forli, Italy}

\email{nagib@illinoistech.edu}

\begin{abstract}
The CICLoPE facility at the University of Bologna in Forli, Italy, is a unique facility that provides fully developed pipe flow up to Reynolds numbers of about $Re_\tau$ of 50,000 with exceptional spatial resolution and stable operating conditions. Measurements obtained over the last two years, on the centerline of the pipe in the fully developed test section, with Pitot probes for the mean velocity in the stream ($\pm 0.2\%$ precision) and hot wires for the intensity of turbulence in the streamwise and the normal stress ( $\pm~5\%$ precision) are reported here and compared to other experimental results and some recent direct numerical simulation (DNS) data.  The comparisons reveal that high Reynolds number conditions are only reached when $Re_\tau$ is at least larger than $10,000$. The centerline von K\`arm\`an constant, $\kappa_{CL}$ for pipe flow is confirmed with high confidence to be $0.44$.  Unlike previously reported measurements on the pipe centerline of the streamwise turbulence intensity and normal stress reported in the literature with various conflicting trends, both quantities are found in CICLoPE to reach a level trend beyond $Re_\tau$ around $25,000$ with values of approximately $2.9\%$ and $0.85$, respectively. Normalized streamwise velocity spectra on the centerline, with high resolution over seven decades of energy, exhibit a clear and extended $-5/3$ inertial range. Skewness and kurtosis of streamwise velocity are found to be independent of Reynolds number even down to $Re_\tau = 5,000$ and equal to $-0.5$ and $3.5$, respectively. \\  

\end{abstract}

\section{Introduction}

The fully developed pipe flow provides several advantages to study turbulence due to its homogeneous boundary conditions and the relative ease to measure the pressure gradient along the pipe and extract the wall shear stress from it.  Although the Superpipe facility at Princeton University allows Reynolds number conditions with $Re_\tau (= u_\tau R / \nu$) about ten times higher than in the CICLoPE facility of the University of Bologna, the spatial resolution accessible in the pipe at CICLoPE is almost an order of magnitude better for similar instrumentation.  One of the simplest but most informative measurements made in a turbulent pipe flow are the centerline data for various quantities as a function of $Re_\tau$. However, data from various facilities in a number of countries with air and water flow in pipes of various sizes using different types of sensors are far from agreeing.  Our aim here is to establish when the conditions representative of high Reynolds number are reached and what are the correct values and trends for several of the streamwise mean and turbulence quantities.

\section{Facility}

The CICLoPE facility consists of a closed-loop ``wind tunnel" with a circular test section $111.5-m$. The facility is located within one of the two 130-m underground tunnels in the former Industrie Caproni, one of the leading aircraft manufacturers in Italy from 1930 to 1943, located in Predappio. These tunnels were excavated beneath the mountains before World
War II to allow aircraft assembly during bombing raids. The closed-loop design ensures stable flow conditions with low turbulence levels. It includes a heat exchanger to maintain temperature control within a range of $\pm 0.1$ degree C and a flow conditioning system comprising honeycomb structures, four screens, a settling chamber, and a convergence with a contraction ratio of 4 is the maximum achievable within the dimensional constraints of the site. Figure 1 provides a drawing of the facility, depicting its key components.

\begin{figure}
      \centering
      \includegraphics[width=0.8\textwidth]{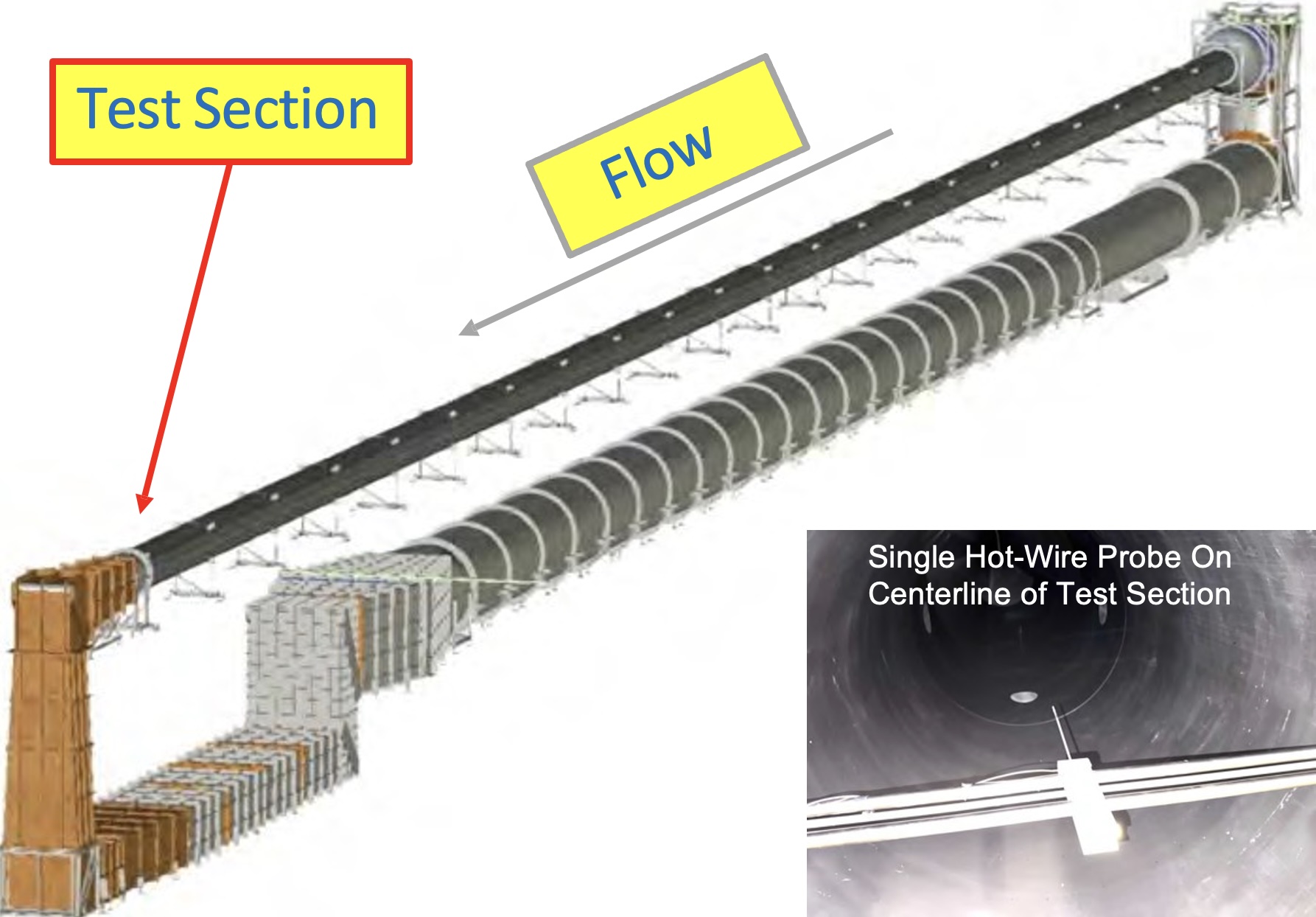}
       \caption{Engineering drawing of CICLoPE facility and photograph of hot-wire probe in test section.}
       \label{fig:fig1}
\end{figure}

The primary component of the flow loop, responsible for $60\%$ of the total pressure losses, is a constant cross-sectional round pipe with an inner diameter of $900$ mm. A photograph of the inside of the test section with a hot-wire probe supported by a traversing mechanism with its sensor on the centerline is part of Figure 1. The pipe measures $111.5$ m in length , resulting in an L/D ratio of approximately 123. It is constructed from twenty-two $5$~m long carbon-fiber sections, followed by a final ``test section" of $1.5$~m length. All elements were produced using filament-winding technology that results in a surface roughness of $k_{rms} < 0.2~\mu$m; i. e., a typical $k^+ < 0.02$ and a diameter tolerance of $900 \pm 0.2$~mm. Each section of the pipe is equipped with four axially spaced static pressure ports and four equally distributed access ports of $150$~mm in diameter. The aluminum access ports are machined to sit flush with the inner surface of the pipe.

The test section connects to the return circuit located one floor below through a circular-to-rectangular shape converter and a series of diffusers and corners with turning vanes. The fan system is located downstream of a cooling heat exchanger and is designed to provide a pressure increase of 6500 Pa at a volume flow rate of $38~m^3/s$, corresponding to a maximum velocity of $60$ m/s in the test section. It consists of two counter-rotating axial fans mounted in series. Each axial fan has two propellers on a shared motor, powered by a dedicated inverter. The diameter of the fan is $1.8$~m, with a total length of $4.2$ m, and the maximum power consumption of the system is $340$ kW. For noise reduction, straight cylindrical sections $20$ m upstream and downstream of the fans are lined with sound-absorbing material. Temperature and humidity are controlled separately for the main tunnel and the laboratory housing the smaller $1.5$~m section, using two external air conditioning systems. Both the airflow of the wind tunnel and the fan motors are liquid cooled via a refrigerating circuit connected to an evaporative tower outside the facility.

\begin{figure}
      \centering
      \includegraphics[width=0.9\textwidth]{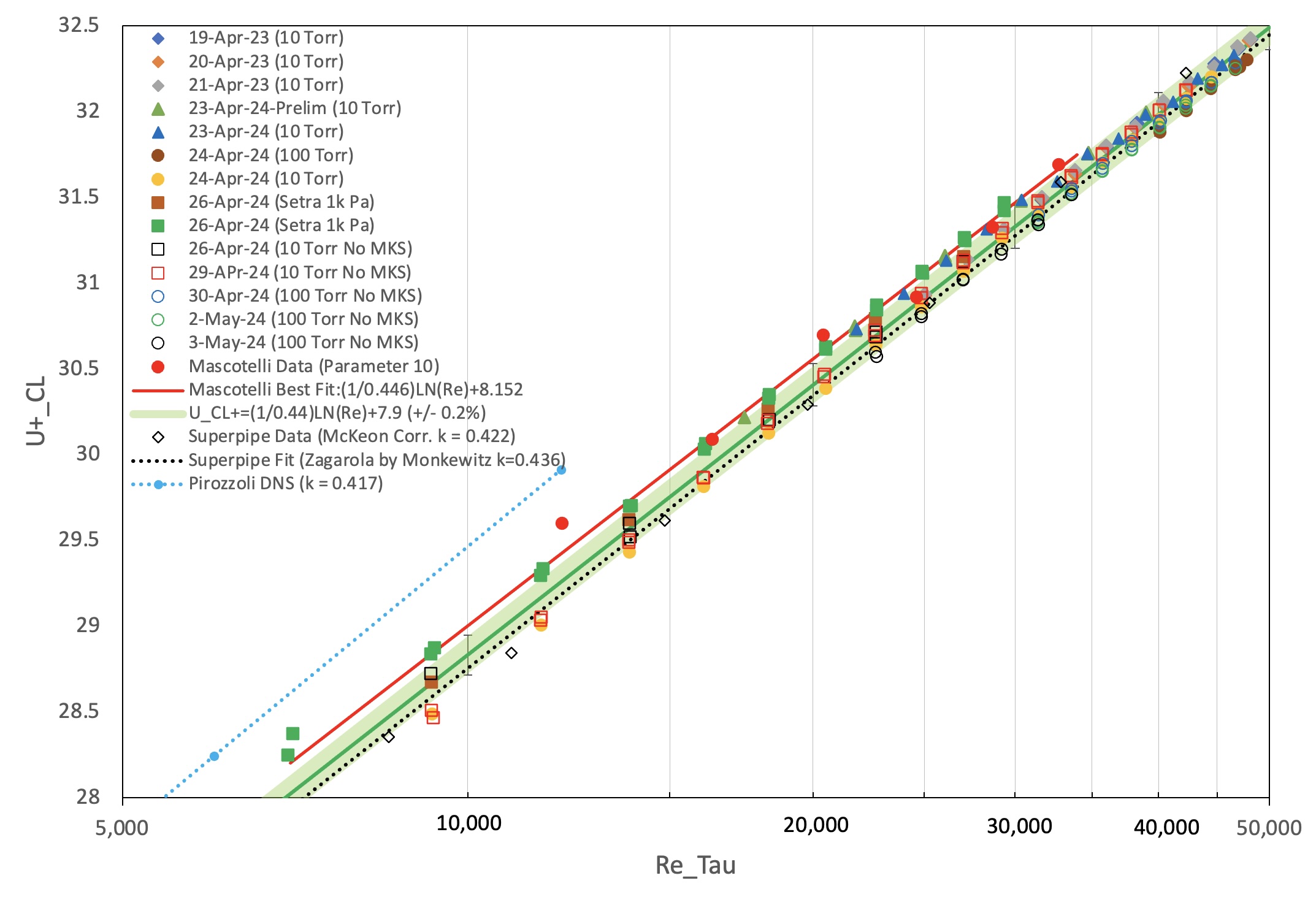}
       \caption{Normalized centerline mean velocity, $U_{CL}^+= U_{CL}/u_\tau$, from CICLoPE by Mascotelli \cite{Lucia} and Lazzarini \cite{Lorenzo}, compared to other data in Superpipe by McKeon et al. \cite{McKeon}, and to DNS results of Pirozzoli \cite{Pirozzoli}.}
       \label{fig:fig2}
\end{figure}

\section{Instrumentation}

The mean velocity on the centerline was measured using a Pitot probe as detailed by Nagib et al. \cite{TSFP}, and the air temperature was continuously monitored, adjusted and recorded to maintain the conditions stated in the above section on the facility. Several pressure transducers were used to optimize the accuracy of the measurements and included Setra (1k Pa), MKS and Baratron (10, 100 and 1000 Torr) transducers.  The Pitot Probe signal was low-pass filtered to prevent aliasing and acquired at 10 kHZ for 30 seconds. The static pressure along the pipe was also measured with these transducers coupled to a Scannivalve or directly with an Initium scanner. The wall shear stress was measured using both the pressure gradient along the fully developed section of the pipe, see Fiorini \cite{Fiorini}, Mascotelli \cite{Lucia}, and also using Oil Film Interferometry (OFI) in the test section, as detailed by Lazzarini \cite{Lorenzo}.  The hot wire measurements were made using a small Dantec single-wire probe. The hot wire signal was low-pass filtered to prevent aliasing and acquired at 60 kHZ for up to 600 seconds.

\section{Results}

The recent CICLoPE results and comparisons with other data are separated into the following two sections on the mean streamwise mean velocity and various representations of the streamwise turbulence velocity.

\subsection{Centerline Mean Velocity}
All current and relevant available data on the normalized centerline mean velocity $U_{CL}^+$ ($= U_{CL}/u_\tau$, where $u_\tau = \sqrt{\tau/\rho}$) are included in Figure 2. During the five years the data was taken, several pressure transducers were used to enhance the accuracy and establish repeatability of the results.  The green shaded area represents the established result of: 

\begin{equation}\label{eq:003}
U_{CL}^+ = (1/0.44) \log{Re_\tau}+ 7.9~(\pm 0.2\%)
\end{equation}

This agrees with the early results of McKeon et al. \cite{McKeon} in Superpipe and those of Nagib et al. \cite{TSFP} in CICLoPE, but with a higher degree of accuracy.

 \begin{figure}
      \centering
      \includegraphics[width=0.8\textwidth]{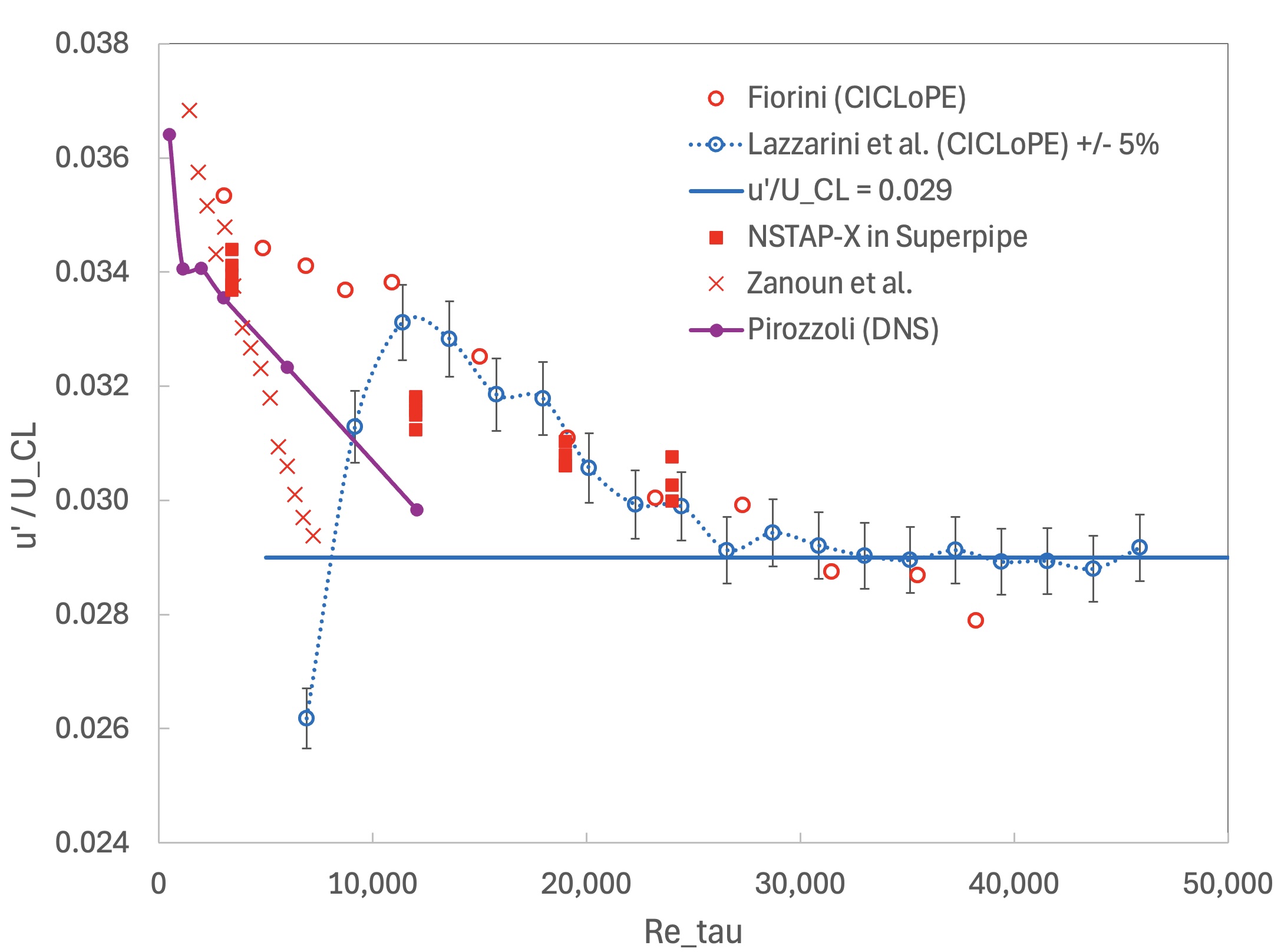}
       \caption{Centerline normalized streamwise turbulence intensity, $u'/U_{CL}$, as function of Reynolds number, $Re_\tau$, from CICLoPE by Fiorini \cite{Fiorini} and Lazzarini \cite{Lorenzo}, and comparisons with DNS data of Pirozzoli \cite{Pirozzoli}, and experimental results by Zanoun et al. \cite{Zanoun24}, and Fu et al. \cite{Fu}.}
       \label{fig:fig3}
\end{figure}

\subsection{Centerline Turbulence}
Figure 3 compares the data for the intensity of the turbulence, $u'/U_{CL}$, by Lazzarini \cite{Lorenzo} at CICLoPE with the earlier results of Fiorini \cite{Fiorini} and recent data of the DNS results by Pirozzoli \cite{Pirozzoli}.  Included are also data from Zanoun et al. \cite{Zanoun24} that have a trend that we cannot explain but that might be related to the values of the measured friction velocity $u_\tau$. Since we did not have access to those values for these results, no further attempt was made to explain them and we were unable to include them in Figure 4. However, we were able to include two other sets of previously reported data in Figure 4 from the Superpipe by Hultmark et al. \cite{Hultmark} and from the pipe installation using water as the working fluid in Japan by Ono et al.\cite{Ono}. Although the low $Re_\tau$ data from the Superpipe are in general agreement with the DNS results of Pirozzoli \cite{Pirozzoli} and the CICLoPE results of Lazzarini \cite{Lorenzo}, their trend towards high Reynolds numbers from the Superpipe was initially unexplained. In addition, the large values of the normal streamwise stress from the Japanese pipe flow cannot be justified.

Figures 3 and 4 indicate that high Reynolds number conditions are only reached when $Re_\tau$ is at least greater than $10,000$.  Unlike previously reported measurements on the pipe centerline of the streamwise turbulence intensity and normal stress reported in the literature with various conflicting trends, both quantities are found in CICLoPE to reach a level trend beyond $Re_\tau$ around $25,000$ with values of approximately $2.9\%$ and $0.85$, respectively.

\begin{figure}
      \centering
      \includegraphics[width=0.8\textwidth]{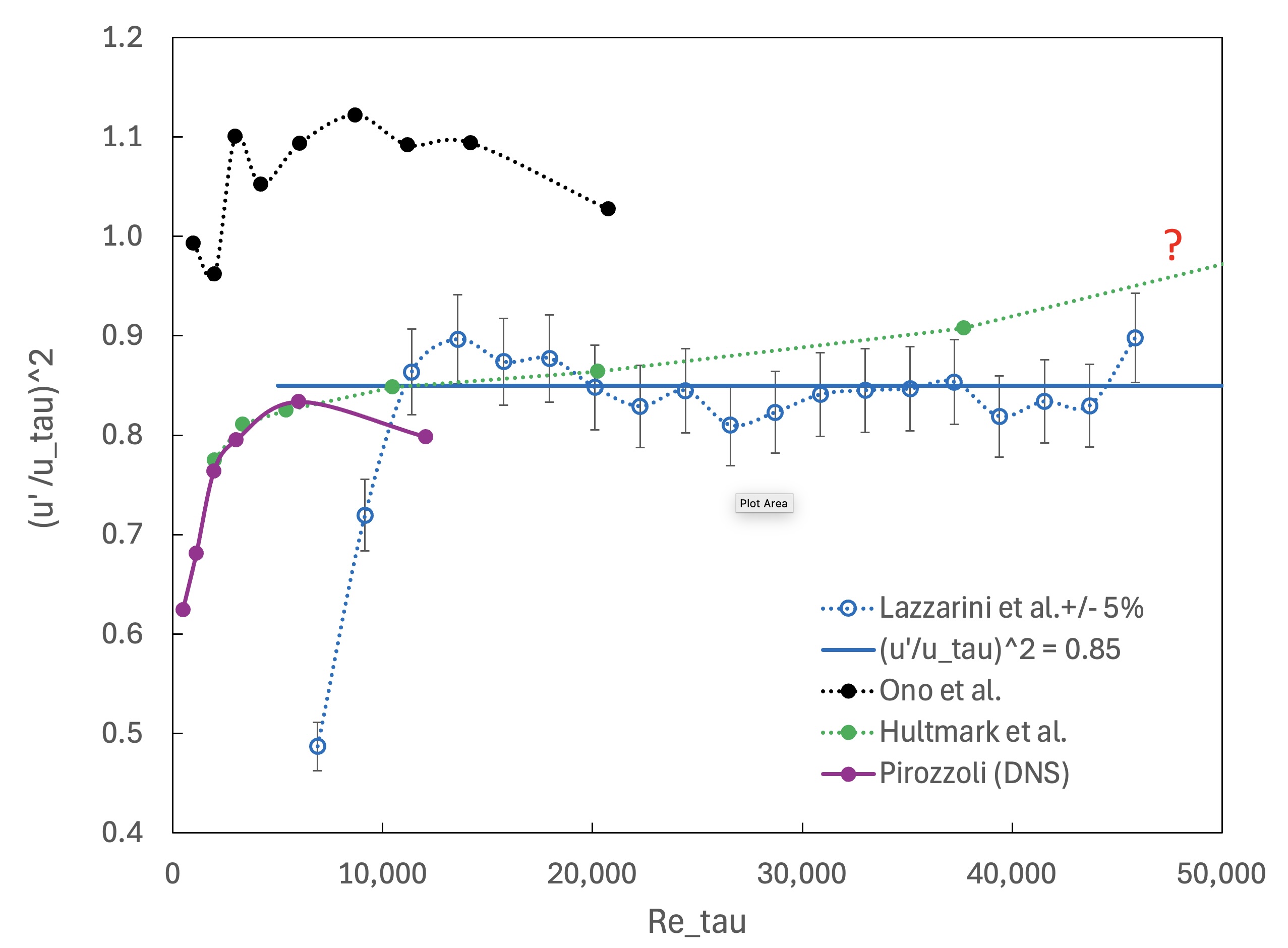}
       \caption{Centerline streamwise normal stress of turbulence, $(u'/u_\tau)^2$, as function of Reynolds number, $Re_\tau$, from CICLoPE by  Lazzarini \cite{Lorenzo}, compared to DNS data from Pirozzoli \cite{Pirozzoli}, and experimental results of  Ono et al. \cite{Ono} and  Hultmark et al. \cite{Hultmark}; see Figure 10 regarding question mark near top right-hand side of this figure.}
       \label{fig:fig4}
\end{figure}

Figure 5 compares the skewness and kurtosis extracted from the time series measured initially at CICLoPE by Fiorini \cite{Fiorini}
 and recently by Lazzarini \cite{Lorenzo}, and establishing their values to be $-0.5$ and $3.5$, respectively.  The excellent comparison extends even to conditions of $Re_\tau$ well below the high Reynolds number trends of Figures 3 and 4, revealing the insensitivity of these two measures to discriminate between transitional conditions and high Reynolds number turbulence.  The normalized stream-wise velocity spectra on the centerline in Figure 6, with high resolution over seven decades of energy, exhibit a clear and extended inertial range with $-5/3$ slope.

\begin{figure} 
      \centering
      \includegraphics[width=0.8\textwidth]{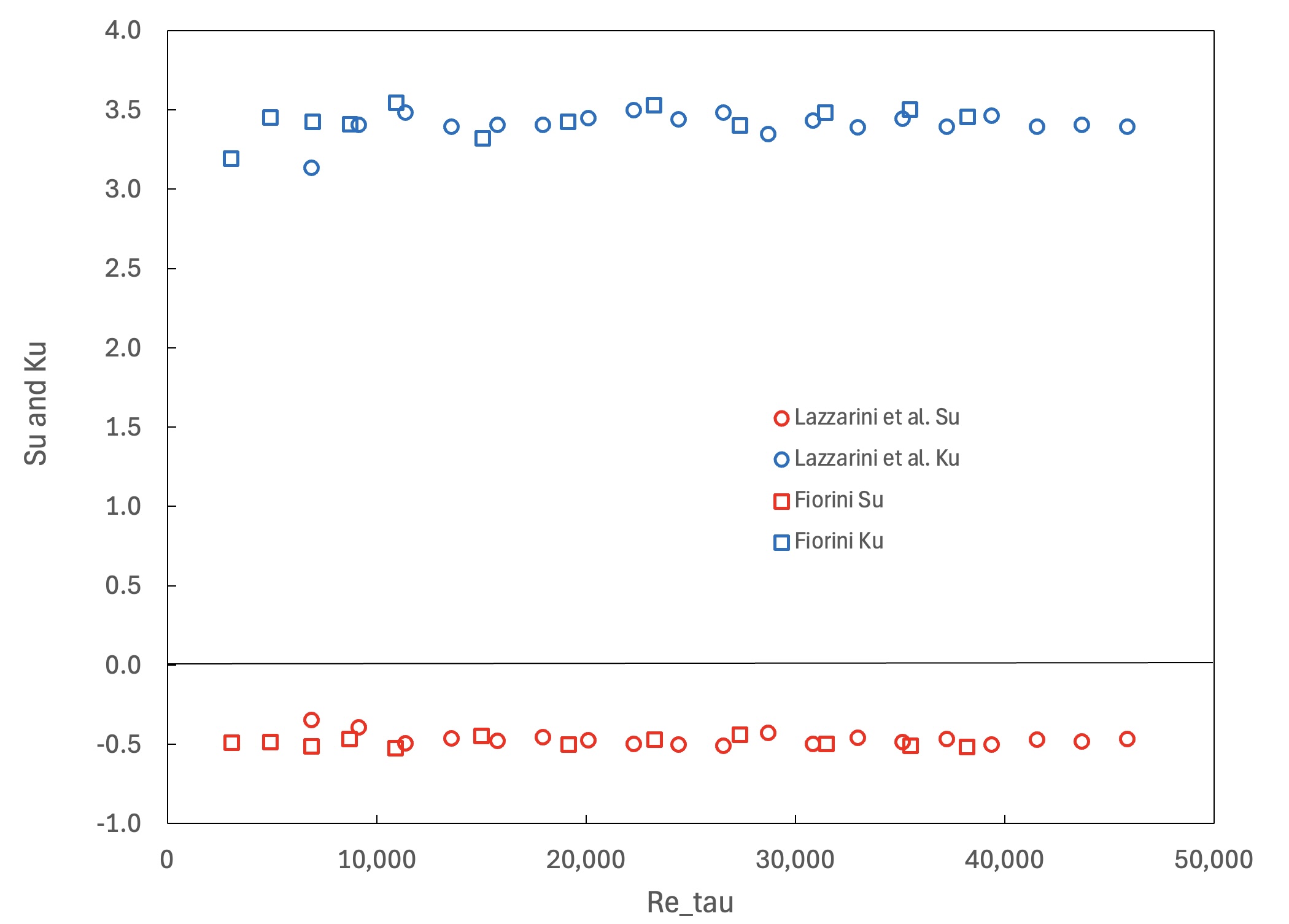}
       \caption{Skewness and kurtosis of streamwise velocity from CICLoPE as function of Reynolds number, $Re_\tau$, by Fiorini \cite{Fiorini} and Lazzarini \cite{Lorenzo}.}
       \label{fig:fig5}
\end{figure}

\begin{figure}
      \centering
      \includegraphics[width=0.98\textwidth]{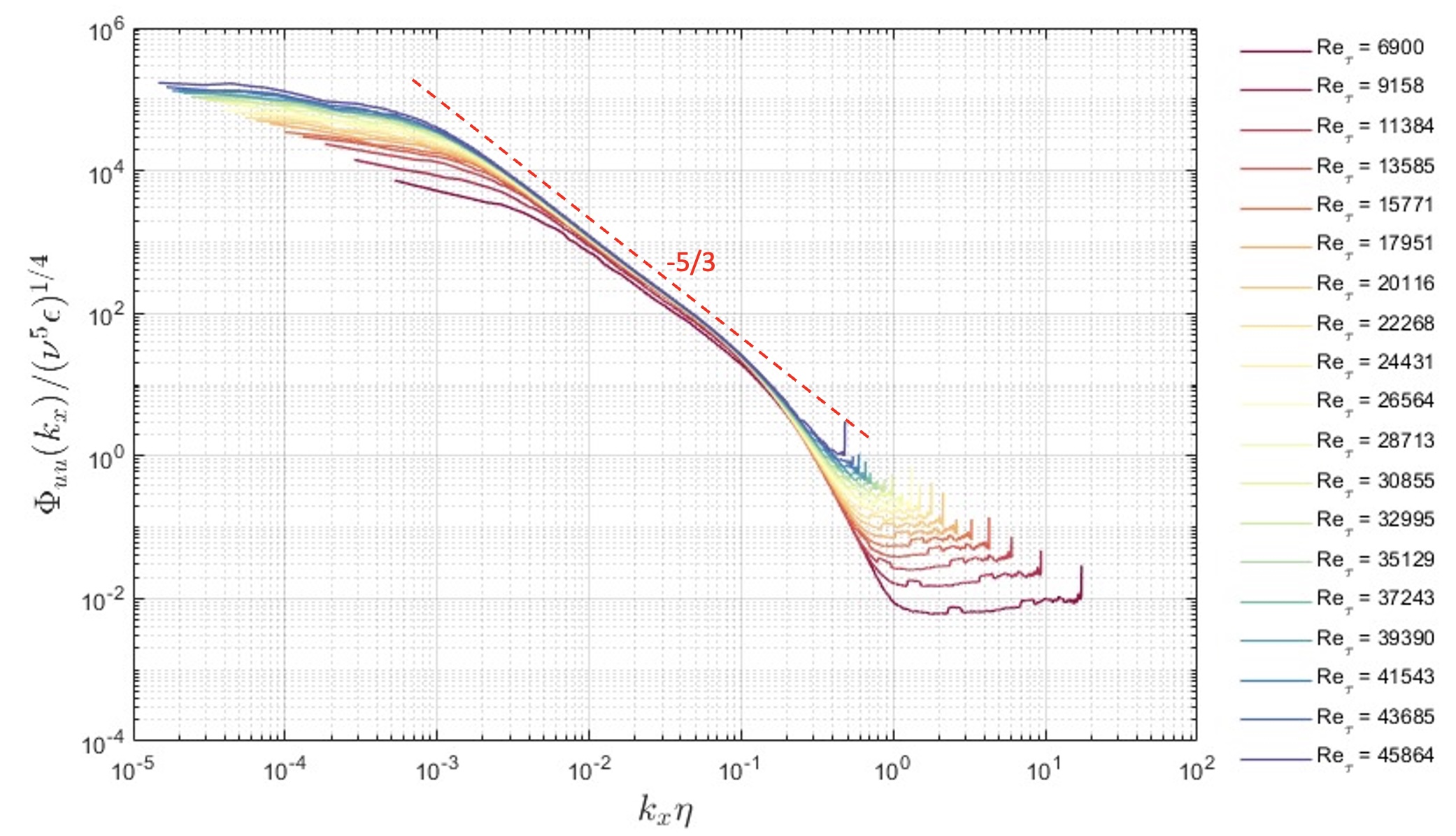}
       \caption{Normalized spectra of streamwise velocity on centerline of pipe for a range of Reynolds numbers, $Re_\tau$, from CICLoPE by Lazzarini \cite{Lorenzo}.}
       \label{fig:fig6}
\end{figure}

\subsection{Processing of unpublished data from Superpipe}
During a visit by Hassan Nagib to Princeton University in spring of 2025, discussions with Marcus Hultmark revealed that some not fully processed data taken by his group and not reported in Fu et al. \cite{FuX} near the centerline of the Superpipe with NSTAP-X probes are available through Matthew K. Fu and Clayton P. Byers.  With their assistance, we processed these data and present several of the results here. Although the data were not acquired exactly on the centerline of the superpipe, the probe position within a few percent of the center of the pipe in the cases we include here. The first comparison with these data is made in Figure 7 showing the ratio of radial and streamwise velocity components of turbulence, $v'/u'$, for various Reynolds numbers compared to recent DNS results of Pirozzoli \cite{Pirozzoli}, and the agreement is generally good.

Processing the NSTAP-X data to extract spectra using FFT revealed interesting trends of additional unexpected low-frequency noise, but only in about half of the cases.  Comparing the noisy cases to ones with less noise at the same Reynolds numbers as in Figure 8 began to reveal the explanation.   Discussions with Matthew K. Fu and Clayton P. Byers. pointed us to the anemometer-related noise reported and analyzed by Hutchins et al. \cite{Nick}. This noise is introduced by the anemometer when the exact characteristics of the probe cable cannot be matched and was found to be intermittent and not appearing for all data runs at the same levels. 

\begin{figure}
      \centering
      \includegraphics[width=0.7\textwidth]{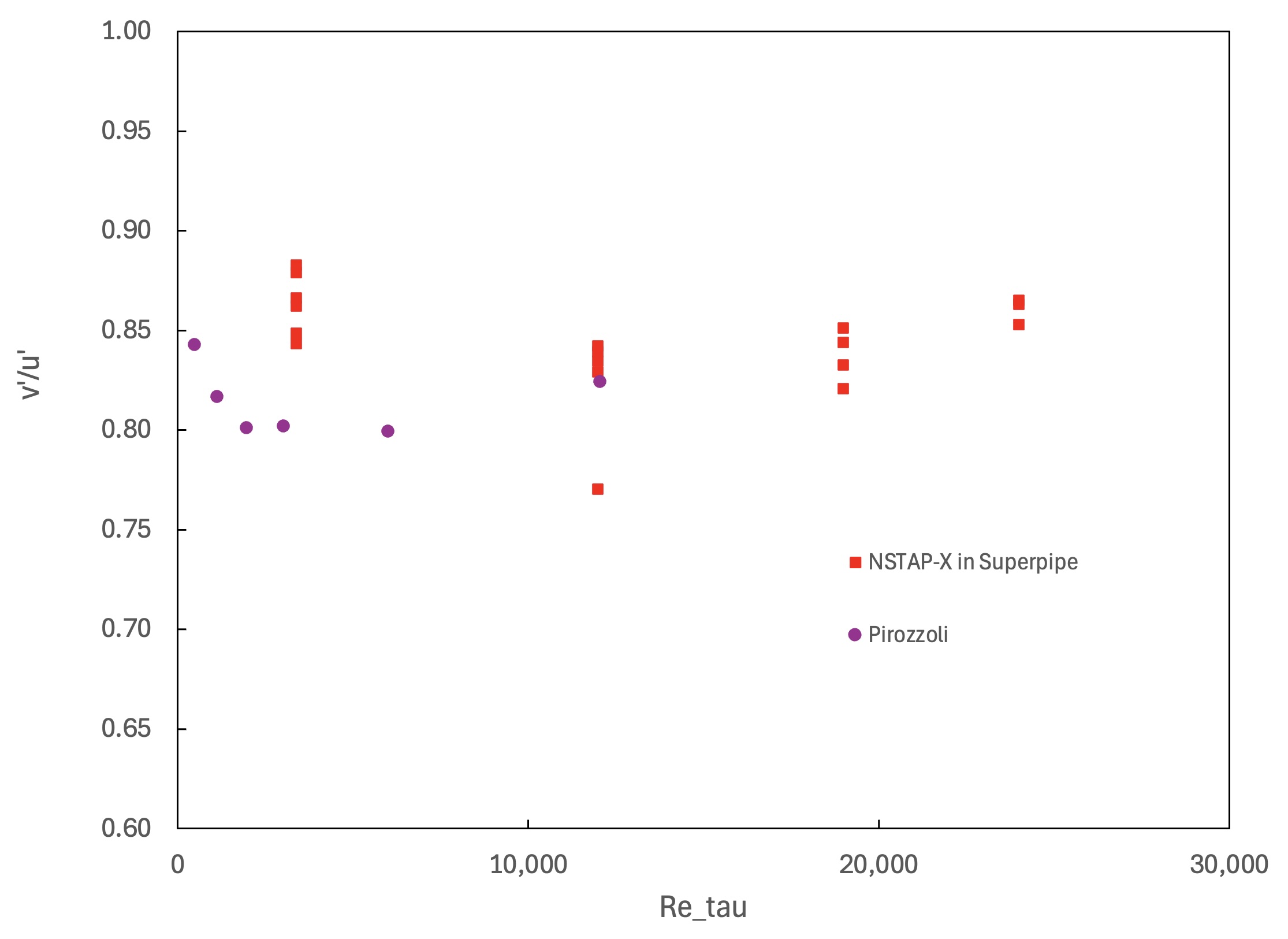}
       \caption{Ratio of turbulence radial and streamwise velocity components, $v'/u'$, measured near centerline of Superpipe by Fu et al. \cite{Fu} compared to DNS of Pirozzoli \cite{Pirozzoli}.}
       \label{fig:fig7}
\end{figure}

\begin{figure}
\centering
\begin{minipage}{.5\textwidth}
  \centering
  \includegraphics[width=.98\linewidth]{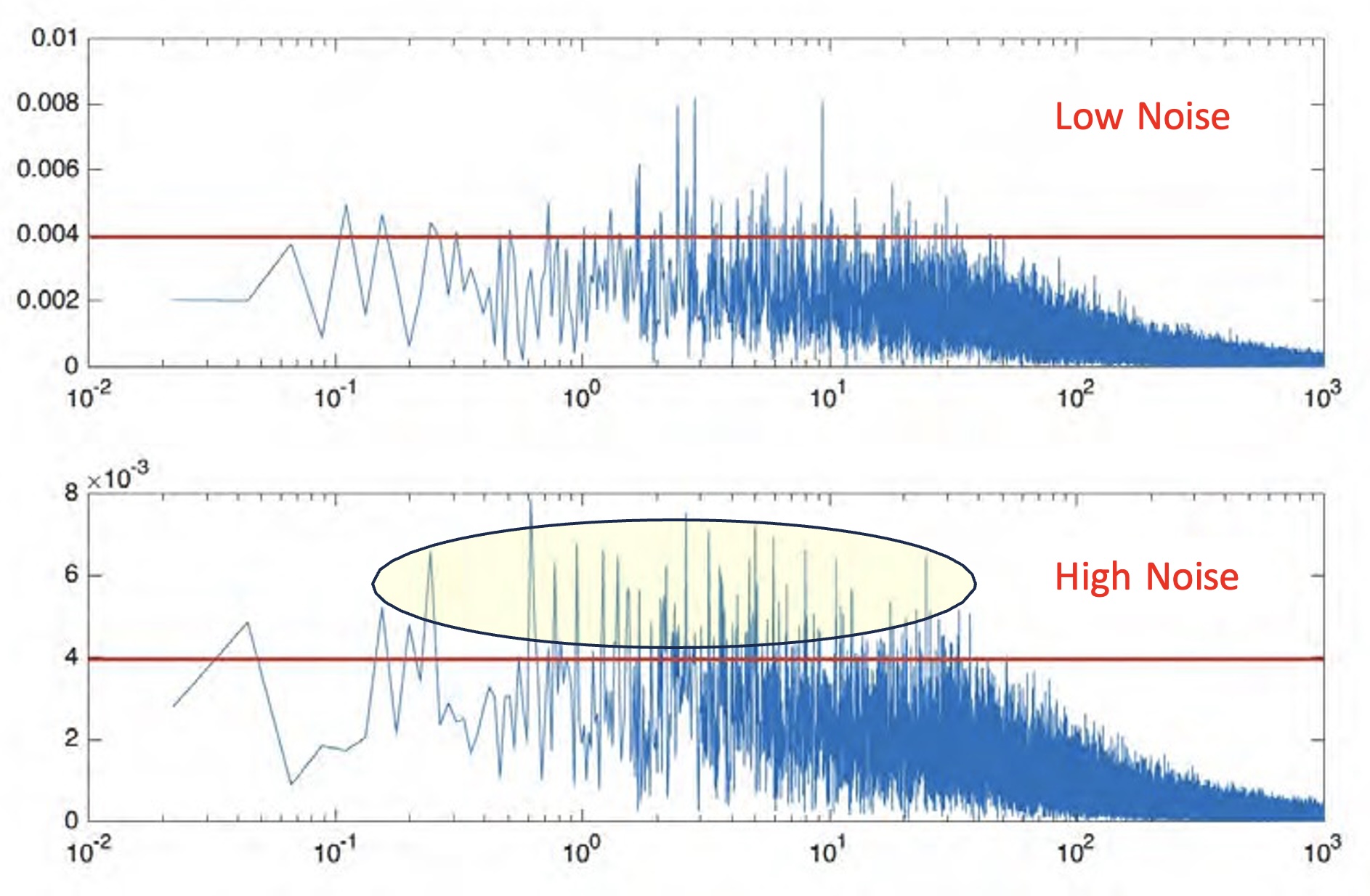}
  \label{fig:test1}
\end{minipage}%
\begin{minipage}{.5\textwidth}
  \centering
  \includegraphics[width=.98\linewidth]{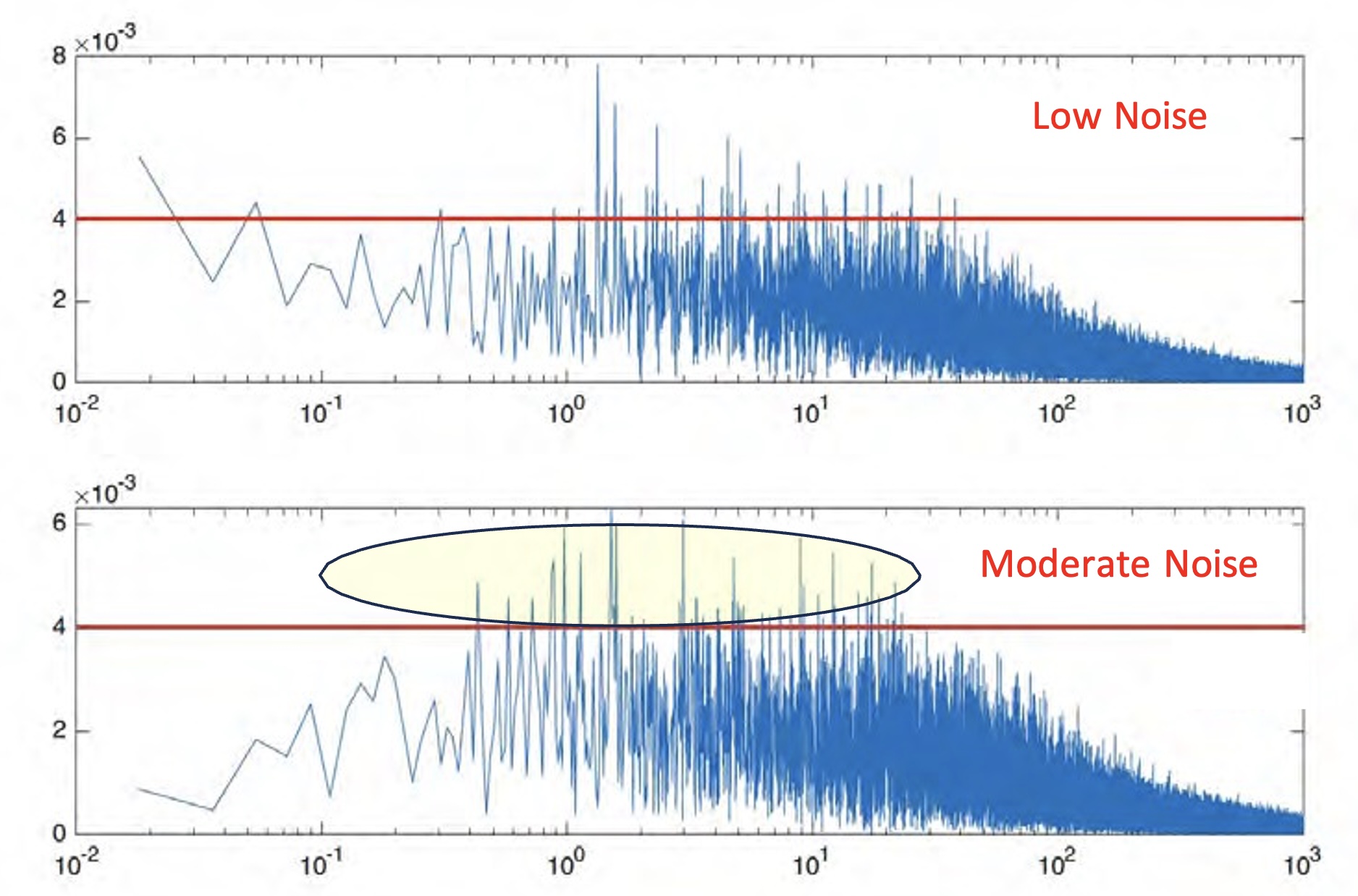}
  \label{fig:test2}
\end{minipage}
   \caption{Spectra of streamwise velocity near centerline of Superpipe measured by NSTAP-X probes, from data provided by Fu et al. \cite{Fu}.}
       \label{fig:fig8}
\end{figure}

\section{Discussion and Conclusions}
A very high degree of care is required to make the type of measurement reported here.  Revealing and explaining the NSTAP results in the Superpipe and the potential influence of anemometer noise on them discussed in last section was an important example. Figure 9 documents an example from CICLoPE, where in one case a post-run calibration of the hot wire was missed and thereby not allowing comparison to the pre-run calibration and confirmation of no drift in the hot wire resistance.

The streamwise normal stress data of Figure 4 are replotted on an extended $Re_\tau$ range in Figure 10. In addition, the processed data from NSTAP-X of Fu et al. \cite{Fu} are included with two types of symbols. When noise was not detected in the spectra, the case is shown with a solid red square. Cases where noise was evident in the spectra are represented with red-open squares.  The agreement between the solid red squares with CICLoPE data is most gratifying and supports our conclusions.  At the same time, the open red squares finally explain the very high $Re_\tau$ data from the Superpipe by Hultmark et al. \cite{Hultmark}.

It is now clear that high Reynolds number conditions are not reached in fully developed pipe flow until $Re_\tau$~exceeds at least $10,000$ and possibly not fully until after $Re_\tau reaches 25,000$. For such high Reynolds number conditions, the streamwise normal stress reaches and remains at a value around 0.85. The mean velocity measurements confirm that $\kappa_{CL} = 0.44$.  Finally, These results establish that CICLoPE is a unique high Reynolds number pipe flow facility with exceptional spatial resolution for the measurements and highly smooth wall surfaces. 

\begin{figure}
      \centering
      \includegraphics[width=0.6\textwidth]{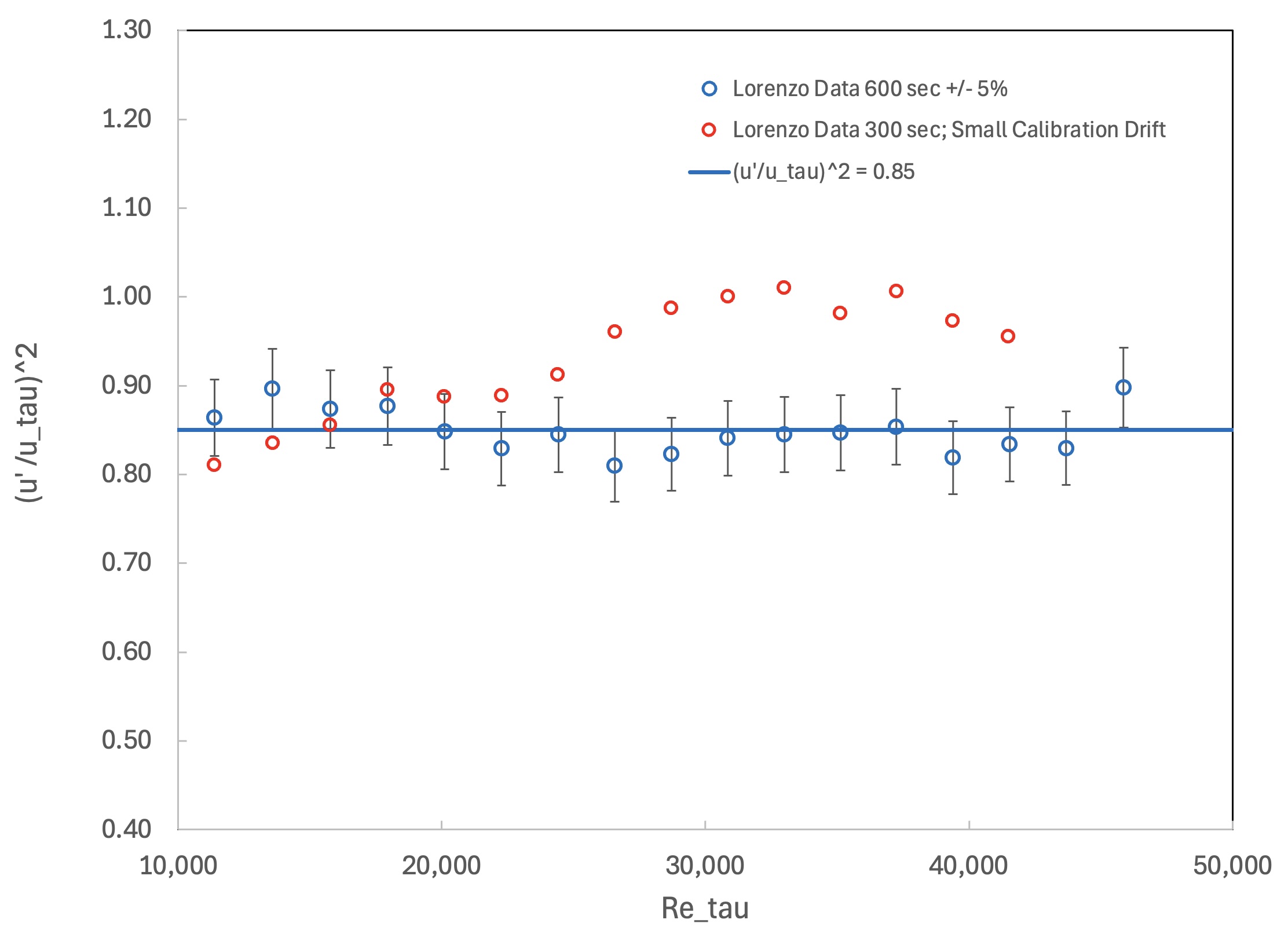}
       \caption{Centerline streamwise normal stress of turbulence, $(u'/u_\tau)^2$, as function of Reynolds number, $Re_\tau$, from CICLoPE demonstrating importance of pre- and pos-calibration of hot wire probes.}
       \label{fig:fig9}
\end{figure}

\begin{figure}
      \centering
      \includegraphics[width=0.9\textwidth]{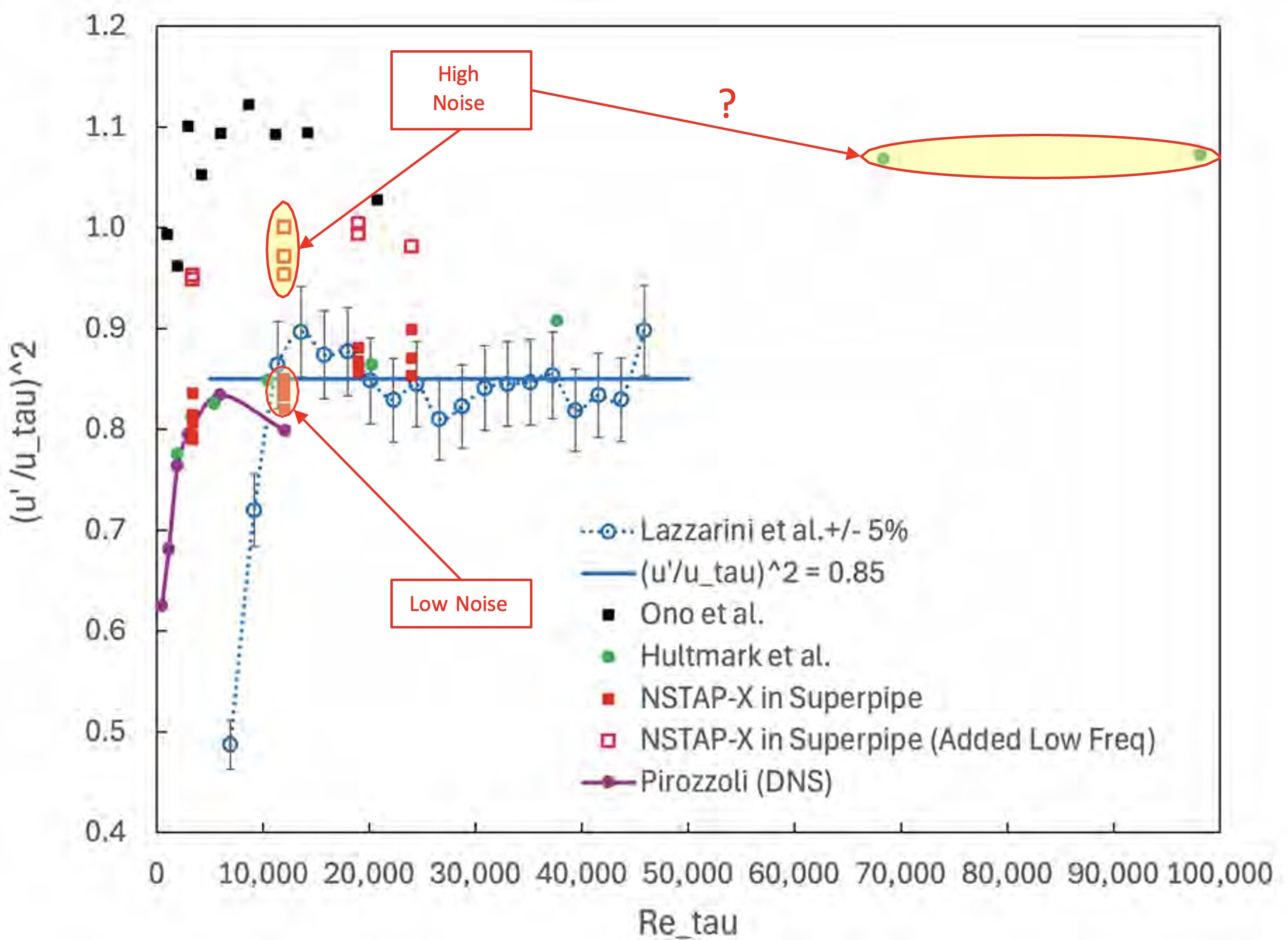}
       \caption{Centerline streamwise normal stress of turbulence, $(u'/u_\tau)^2$, as function of Reynolds number, $Re_\tau$, from CICLoPE by  Lazzarini \cite{Lorenzo}, compared to results of DNS by Pirozzoli \cite{Pirozzoli}, and experiments of  Ono et al. \cite{Ono},  Hultmark et al. \cite{Hultmark}, and multiple data series under the same Reynolds number, $Re_\tau$, provided to us by Fu et al. \cite{Fu}.}.
       \label{fig:fig10}
\end{figure}

\section{Acknowledgment}
The authors thank Matthew K. Fu, Clayton P. Byers, and Marcus Hultmark for sharing the NSTAP-X data with us and for the insight they provided on the source of the electronic noise in some of the data.

\bibliography{main}

\end{document}